\documentclass{article}    

\usepackage{graphicx}
\usepackage{dcolumn}
\usepackage{bm}

\begin{document}           

\title{Finite bounded expanding white hole universe without dark matter}  

\author{\textbf{John G. Hartnett}\\
School of Physics, the University of Western Australia,\\
 35 Stirling Hwy, Crawley 6009 WA Australia\\
\textit{john@physics.uwa.edu.au}}

\maketitle  

\begin{abstract}
The solution of Einstein's field equations in Cosmological General Relativity (CGR), where the Galaxy is at the center of a finite yet bounded spherically symmetrical isotropic gravitational field, is identical with the unbounded solution. This leads to the conclusion that the Universe may be viewed as a finite expanding white hole. The fact that CGR has been successful in describing the distance modulus verses redshift data of the high-redshift type Ia supernovae means that the data cannot distinguish between unbounded models and those with finite bounded radii of at least $c \tau$. Also it is shown that the Universe is spatially flat at the current epoch and has been at all past epochs where it was matter dominated.
\end{abstract}


\section{\label{sec:Intro}Introduction}

In an interview with Scientific American George Ellis once said \cite{Gibbs1995}

``People need to be aware that there is a range of models that could explain the observations, \ldots For instance, I can construct you a spherically symmetrical universe with Earth at its center, and you cannot disprove it based on observations. \ldots You can only exclude it on philosophical grounds. In my view there is absolutely nothing wrong in that. What I want to bring into the open is the fact that we are using philosophical criteria in choosing our
models. A lot of cosmology tries to hide that.''

This paper proposes a model where the Galaxy is at the center of a spherically symmetrical finite bounded universe. It contends that fits to the magnitude-redshift data of the high-$z$ type Ia supernovae (SNe Ia) \cite{Knop2003, Riess2004, Astier2005}, are also consistent with this model. That is, providing that the radius of the Universe (a spherically symmetrical matter distribution) is at least $c\tau$ where $c$ is the speed of light and $\tau \approx 4.28 \times 10^{17}$ s (or $13.54 \; Gyr$).\cite{Oliveira2006} Here $\tau$ is the Hubble-Carmeli time constant, or the inverse of the Hubble constant evaluated in the limits of zero gravity and zero distance.

This model is based on the Cosmological General Relativity (CGR) theory \cite{Carmeli2002a} but explores the motion of particles in a central potential.  In this case the central potential is the result of the Galaxy being situated at the center of a spherically symmetrical isotropic distribution comprising all matter in the Universe. 

This paper is preceded by Hartnett \cite{Hartnett2005} that forms the basis of the work presented here. Also Oliveira and Hartnett \cite{Oliveira2006} progressed the work by developing a density function for higher redshifts. Those paper assumed the unbounded model. The reader should be familiar with Hartnett \cite{Hartnett2005} at least before reading this. 
 
\subsection{\label{sec:CGR}Cosmological General Relativity}

The metric \cite{Behar2000, Carmeli1996, Carmeli2002a} used by Carmeli (in CGR) in a generally covariant theory extends the number of dimensions of the Universe by the addition of a new dimension -- the radial velocity of the galaxies in the Hubble flow. The Hubble law is assumed as a fundamental axiom for the Universe and the galaxies are distributed accordingly. The underlying mechanism is that the substance of which space is built, the vacuum, is uniformly expanding in all directions and galaxies, as tracers, are fixed to space and therefore the redshifts of distant first ranked galaxies quantify the speed of the expansion.

In determining the large scale structure of the Universe the usual time dimension is neglected ($dt = 0$) as observations are taken over such a short time period compared to the motion of the galaxies in the expansion. It is like taking a still snap shot of the Universe and therefore only four co-ordinates $x^{\mu}=(x^{1},x^{2},x^{3},x^{4})=(r,\theta,\phi,\tau v)$ are used -- three of space and one of velocity. The parameter $\tau$, the Hubble-Carmeli constant, is a universal constant for all observers. 

Here the CGR theory is considered using a Riemannian four-dimensional presentation of gravitation in which the coordinates are those of Hubble, i.e. distance and velocity. This results in a phase space equation where the observables are redshift and distance. The latter may be determined from the high-redshift type Ia supernova observations. 

\subsection{\label{sec:phasespace}Phase space equation}

The line element in CGR \cite{Carmeli2002b}
\begin{equation} \label{eqn:lineelement}
ds^2= \tau^{2}dv^{2}-e^{\xi}dr^{2}-R^2(d\theta^2+ sin^2\theta d\phi^2),
\end{equation}
represents a spherically symmetrical isotropic universe, that is not necessarily homogeneous. 

It is fundamental to the theory that $ds=0$. In the case of Cosmological Special Relativity (see chap.2 of \cite{Carmeli2002a}), which is very useful pedagogically, we can write the line element as
\begin{equation} \label{eqn:CSRlineelement}
ds^2= \tau^{2}dv^{2}-dr^{2},
\end{equation}
ignoring $\theta$ and $\phi$ co-ordinates for the moment. By equating $ds=0$ it follows from (\ref{eqn:CSRlineelement}) that $\tau dv = dr$ assuming the positive sign for an expanding universe. This is then the Hubble law in the small $v$ limit. Hence, in general, this theory requires that $ds =0$. 

Using spherical coordinates ($r,\theta, \phi$) and the isotropy condition $d\theta = d\phi = 0 $ in (\ref{eqn:lineelement}) then $dr$ represents the radial co-ordinate distance to the source and it follows from (\ref{eqn:lineelement}) that
\begin{equation} \label{eqn:4Dmetric}
\tau^{2}dv^{2}-e^{\xi}dr^{2}=0,
\end{equation}
where $\xi$ is a function of $v$ and $r$ alone. This results in
\begin{equation} \label{eqn:4Dmetricderiv}
\frac{dr} {dv} = \tau e^{-\xi/2},
\end{equation}
where the positive sign has been chosen for an expanding universe.

\section{\label{sec:Centralsoln}Solution in central potential}

Carmeli found a solution to his field equations, modified from Einstein's, (see \cite{Hartnett2005} and \cite{Behar2000,Carmeli2002a,Carmeli2002b}) which is of the form
\begin{equation} \label{eqn:soln1field2}
e^{\xi}= \frac{{R'^2}}{1 + f(r)}
\end{equation}
with $R' = 1 $, which must be positive. From the field equations and (\ref{eqn:soln1field2}) we get a differential equation
\begin{equation} \label{eqn:soln1field3}
f' + \frac{f}{r} = - \kappa  \tau^{2} \rho_{eff} r,
\end{equation}
where $f(r)$ is function of $r$ and satisfies the condition $f(r) + 1 > 0$. The prime is the derivative with respect to $r$. Here $\kappa = 8 \pi G /c^{2} \tau^{2}$ and $\rho_{eff} = \rho -\rho_{c}$ where $\rho$  is the averaged matter density of the Universe and $\rho_{c}=3/8 \pi G \tau^{2}$ is the critical density. 

The solution of (\ref{eqn:soln1field3}), $f(r)$, is the sum of the solution ($2GM/c^{2}r$) to the homogeneous equation  and a particular solution (-$\frac{\kappa}{3}  \tau^{2} \rho_{eff} r^{2} $) to the inhomogeneous equation. In \cite{Carmeli2002a} Carmeli discarded the homogeneous solution  saying it was not relevant to the Universe, but the solution of a particle at the origin of coordinates, or in other words, in a central potential.

Now suppose we model the Universe as a ball of dust of radius $\Delta$ with us, the observer, at the center of that ball. In this case the gravitational potential written in spherical coordinates that satisfies Poisson's equation in the Newtonian approximation is
\begin{equation} \label{eqn:potential1}
\Phi(r) =-\frac{GM}{r}
\end{equation}
for the vacuum solution, but inside an isotropic matter distribution
\begin{eqnarray} \label{eqn:potential2}
\Phi(r)&&=-G \left (\frac{4\pi \rho}{r} \int_{0}^{r} r'^{2} dr' + 4\pi \rho \int_{r}^{\Delta} r' dr'\right) \nonumber \\
&&= \frac{2}{3} G \pi \rho r^{2} - 2 G \pi \rho \Delta^{2},
\end{eqnarray}
where it is assumed the matter density $\rho$ is uniform throughout the Universe. At the origin ($r = 0$) $\Phi(0) = - 2 G \pi \rho_{m} \Delta^{2}$, where $\rho = \rho_{m}$ the matter density at the present epoch. In general $\rho$ depends on epoch. Because we are considering no time development $\rho$ is only a function of redshift $z$ and $\rho_{m}$ can be considered constant.

From (\ref{eqn:potential2}) it is clear to see that by considering a finite distribution of matter of radial extent $\Delta$, it has the effect of adding a constant to $f(r)$ that is consistent with the constant $2 G \pi \rho \Delta^{2}$ in (\ref{eqn:potential2}), where $f(r)$ is now identified with $-4\Phi/c^{2}$. 

Equation (\ref{eqn:soln1field2}) is essentially Carmeli's equation A.19, the solution to his equation A.17 from p.122 of \cite{Carmeli2002a}. More generally (\ref{eqn:soln1field2})  can be written as
\begin{equation} \label{eqn:soln1fieldgeneral}
e^{\xi}= \frac{{R'^{2}}}{1 + f(r) - K},
\end{equation}
where $K$ is a constant.  This is the most general form of the solution of Carmeli's equation A.17.
So by substituting (\ref{eqn:soln1fieldgeneral}) into Carmeli's A.18, A.21 becomes instead
\begin{equation} \label{eqn:generalfield2}
\frac{1}{R R'}(2 \dot{R} \dot{R'} - f') + \frac{1}{R^{2}}(\dot{R}^{2}- f + K) = \kappa \tau^2 \rho_{eff}.
\end{equation}

Therefore (\ref{eqn:soln1fieldgeneral}) is also a valid solution of the Einstein field equations (A.12 - A.18 \cite{Carmeli2002a}) in this model. Making the assignment $R = r$ in (\ref{eqn:generalfield2}) yields a more general version of (\ref{eqn:soln1field3}), that is,
\begin{equation} \label{eqn:soln1fieldgen2}
f' + \frac{f -  K}{r} = - \kappa  \tau^{2} \rho_{eff} r.
\end{equation}

The solution of (\ref{eqn:soln1fieldgen2}) is then
\begin{equation} \label{eqn:fvalue1}
f(r) = -\frac{1}{3}\kappa \tau^{2} \rho_{eff} r^{2}+ K.
\end{equation}
From  a comparison with (\ref{eqn:potential2}) it would seem that the constant $K$ takes the form $K = 8 \pi G \rho_{eff}(0) \Delta^{2}/c^{2}$. It is independent of $r$ and describes a non-zero gravitational potential of a finite universe measured at the origin of coordinates. There is some ambiguity however as to which density to use in Carmelian cosmology since it is not the same as Newtonian theory. Here $\rho_{eff}$ is used and evaluated at $r = 0$.

In the above Carmelian theory it initially assumed that the Universe has expanded over time and at any given epoch it has an averaged density $\rho$, and hence $\rho_{eff}$. The solution of the field equations has been sought on this basis. However because the Carmeli metric is solved in an instant of time (on a cosmological scale) any time dependence is neglected. In fact, the general time dependent solution has not yet been found. But since we observe the expanding Universe with the coordinates of Hubble at each epoch (or redshift $z$) we see the Universe with a different density $\rho(z)$ and an effective density  $\rho_{eff} (z)$. Carmeli arrived at his solution with the constant density assumption. I have made the implicit assumption that the solution is also valid if we allow the density to vary as a function of redshift, as is expected with expansion. 

Now it follows from (\ref{eqn:4Dmetricderiv}), (\ref{eqn:soln1fieldgeneral}) and (\ref{eqn:fvalue1}) that 
\begin{equation} \label{eqn:phasespacederiv2}
\frac{dr}{dv}= \tau \sqrt{1+ \left(\frac{1-\Omega}{c^{2} \tau^{2}}\right) r^{2} } ,
\end{equation}
where $\Omega = \rho/\rho_{c}$. This compares with the solution when the central potential is neglected (i.e. $\Delta \rightarrow 0$). In fact, the result is identical as we would expect in a universe where the Hubble law is universally true.

Therefore (\ref{eqn:phasespacederiv2}) may be integrated exactly and yields the same result as Carmeli,
\begin{equation} \label{eqn:phasespacesolnnatural}
\frac {r} {c \tau}= \frac {\sinh (\frac{v}{c} \sqrt{1-\Omega})} {\sqrt{1-\Omega}}.
\end{equation}

Since observations in the distant cosmos are always in terms of redshift, $z$, we write (\ref{eqn:phasespacesolnnatural}) as a function of redshift where $r$ is expressed in units of $c \tau$ and $v/c = ((1+z)^2-1)/((1+z)^2+1)$ from the relativistic Doppler formula. The latter is appropriate since this is a velocity dimension. 

What is important to note though is that regardless of the geometry of the Universe, provided it is spherically symmetrical and isotropic on the large scale, (\ref{eqn:phasespacesolnnatural}) is identical to that we would get where the Universe has a unique center, with one difference which is explored in the following section. For an isotropic universe without a unique center, one can have an arbitrary number of centers. However if we are currently in a universe where the Galaxy is at the center of the local isotropy distribution this means the Universe we see must be very large and we are currently limited from seeing into an adjacent region with a different isotropy center. 

\section{\label{sec:GravitationalPotential}Gravitational Redshift}

In Hartnett \cite{Hartnett2005} the geometry in the model is the usual unbounded type, as found in an infinite universe, for example. In a finite bounded universe, an additional effect may result from the photons being received from the distant sources. The gravitational redshift ($z_{grav}$) resulting from the Galaxy sitting at the unique center of a finite spherically symmetrical matter distribution must be considered. In this case we need to consider the difference in gravitational potential between the points of emission and reception of a photon. Now the 00th metric component, the time part of the 5D metric of coordinates $x^k=t, r, \theta, \phi, v$ ($k = 0-4$), is required but it has never been determined for the cosmos in the Carmelian theory. In general relativity we would relate it by $g_{00} = 1-4\Phi/c^2$ where $-4\Phi$ is the gravitational potential. The factor 4 and minus sign arise from the Carmelian theory when (\ref{eqn:fvalue1}) and (\ref{eqn:potential2}) are compared. So the question must be answered, ``What is $g_{00}$ metric component for the large scale structure of the universe in CGR?''

First note from (\ref{eqn:soln1field2}) and (\ref{eqn:soln1field3}) the $g_{11}$ metric component (considered in an unbounded universe for the moment)
\begin{equation} \label{eqn:g11}
g_{11} = -\left(1+\frac{1-\Omega}{c^2\tau^2}r^2\right)^{-1}
\end{equation}
in CGR we can write a scale radius 
\begin{equation} \label{eqn:scaleradius}
R = \frac{c \tau}{\sqrt{|1-\Omega|}}. 
\end{equation}
Hence we can define an energy density from the curvature
\begin{equation} \label{eqn:curvaturedensity}
\Omega_K = \frac{c^2}{h^2 R^2} = \frac{c^2 \tau^2}{R^2},
\end{equation}
which, when we use (\ref{eqn:scaleradius}), becomes 
\begin{equation} \label{eqn:curvaturedensity2}
\Omega_K = 1-\Omega.
\end{equation}
This quantifies the energy in the curved \textit{spacevelocity}.

In the FRW theory the energy density of the cosmological constant is defined $\rho_{\Lambda} = \Lambda/8 \pi G$ hence
\begin{equation} \label{eqn:lambdadensityFRW}
\Omega_{\Lambda} = \frac{\Lambda}{3 H^2_0}.
\end{equation}
Even though the cosmological constant is not explicitly used in CGR, it follows from the definition of the critical density that
\begin{equation} \label{eqn:criticaldensityCGR}
\rho_c = \frac{3}{8 \pi G \tau^2} = \frac{\Lambda}{8 \pi G}, 
\end{equation}
when the cosmological constant $\Lambda$ is identified with $3/\tau^2$. Therefore in CGR it follows that
\begin{equation} \label{eqn:lambdadensityCGR}
\Omega_{\Lambda} = \frac{\Lambda}{3 h^2} = \Lambda\left(\frac{\tau^2}{3}\right) = 1.
\end{equation}
This means that in CGR the vacuum energy $\rho_{vac} = \Lambda/8\pi G$ is encoded in the metric via the critical density since $\rho_{eff} = \rho - \rho_c$ principally defines the physics. So $\Omega_{\Lambda} = 1$ identically and at all epochs of time. (The determination of $\Omega_{\Lambda}$ in \cite{Hartnett2005} was flawed due to an incorrect definition.) Also we can relate $\Omega_{\Lambda}$ to the curvature density by
\begin{equation} \label{eqn:curvaturedadensity3}
\Omega_K = \Omega_{\Lambda} - \Omega,
\end{equation}
which becomes
\begin{equation} \label{eqn:curvaturedadensity4}
\Omega_k = \Omega_{\Lambda} - \Omega_m,
\end{equation}
at the present epoch ($z \approx 0$). Here $\Omega=\Omega_m(1+z)^3$ and hence $\Omega_K \rightarrow \Omega_k$ as $z \rightarrow 0$.

Finally we can write for the total energy density, the sum of the matter density and the curvature density,
\begin{equation} \label{eqn:totaldensity}
\Omega_t = \Omega +  \Omega_K = \Omega +  1 - \Omega = 1,
\end{equation}
which means the present epoch value is trivially
\begin{equation} \label{eqn:totaldensity0}
\Omega_0 = \Omega_m +  \Omega_k = \Omega_m +  1 - \Omega_m = 1.
\end{equation}
This means that the 3D spatial part of the Universe is always flat as it expands. This explains why we live in a universe that we observe to be identically geometrically spatially flat. The curvature is due to the velocity dimension. Only at some past epoch, in a radiation dominated universe, with radiation energy density $\Omega_R (1+z)^4$, would the total mass/energy density depart from unity. 

Now considering a finite bounded universe, from (\ref{eqn:fvalue1}), using $\Omega = \rho/\rho_{c}$, I therefore write $g_{00}$ as
\begin{equation} \label{eqn:g00}
g_{00}(r) = 1 + (1-\Omega_t) r^2 + 3(\Omega_t-1) \Delta^2,
\end{equation}
where $r$ and $\Delta$ are expressed in units of $c \tau$. Equation (\ref{eqn:g00}) follows from $g_{00} = 1-4\Phi/c^2$  where $\Phi$ is taken from the gravitational potential but with effective density, which in turn involves the total energy density because we are now considering \textit{spacetime}.

Clearly from (\ref{eqn:totaldensity}) it follows that $g_{00}(r) = 1$ regardless of epoch.
Thus from the usual relativistic expression 
\begin{equation} \label{eqn:gravredshift}
1 + z_{grav} = \sqrt{\frac{g_{00}(0)}{g_{00}(r)}} =1,
\end{equation}
and the gravitational redshift is zero regardless of epoch. As expected if the emission and reception of a photon both occur in flat space then we'd expect no gravitational effects.

In an unbounded universe, though no gravitational effects need be considered, the result $g_{00} = 1$ is also the same. Therefore we can write down the full 5D line element for CGR in any dynamic spherically symmetrical isotropic universe,
\begin{equation} \label{eqn:linelement5D}
ds^2 = c^2dt^2 -\left(1+\frac{1-\Omega}{c^2\tau^2}r^2\right)^{-1}dr^2 +\tau^2 dv^2.
\end{equation}
The $\theta$ and $\phi$ coordinates do not appear due to the isotropy condition $d\theta = d\phi=0$. Due to the Hubble law the 2nd and 3rd terms sum to zero leaving $dt = ds/c$, the proper time. Clocks, co-moving with the galaxies in the Hubble expansion, would measure the same proper time.

Since it follows from (\ref{eqn:g00}) that $g_{00}(r) = 1$ regardless of epoch,  $g_{00}(r)$ is not sensitive to  any value of $\Delta$. This means the above analysis is true regardless of whether the universe is bounded or unbounded. The observations cannot distinguish. In an unbounded or bounded universe of any type no gravitational redshift (due to cosmological causes) in light from distant source galaxies would be observed. 

However inside the Galaxy we expect the matter density to be much higher than critical, ie $\Omega_{galaxy} \gg 1$ and the total mass/energy density can be written
\begin{equation} \label{eqn:totaldensitygal}
\Omega_0 |_{galaxy} = \Omega_{galaxy} +  \Omega_k \approx \Omega_{galaxy},
\end{equation}
because $\Omega_k \approx 1$, since it is cosmologically determined. Therefore this explains why the galaxy matter density only is appropriate when considering the Poisson equation for galaxies.\cite{Hartnett2005b} 

As a result inside a galaxy we can write 
\begin{equation} \label{eqn:g00Galaxy}
g_{00}(r) = 1 + \Omega_K \frac{r^2}{c^2\tau^2} + \Omega_{galaxy} \frac{r^2}{c^2\tau^2} ,
\end{equation}
in terms of densities at some past epoch. Depending on the mass density of the galaxy, or cluster of galaxies, the value of $g_{00}$ here changes. As we approach larger and larger structures it mass density approaches that of the Universe as a whole and $g_{00} \rightarrow 1$ as we approach the largest scales of the Universe. Galaxies in the cosmos then act only as local perturbations but have no effect on $\Omega_K$. That depends only on the average mass density of the whole Universe, which depends on epoch ($z$).

Equation (\ref{eqn:g00Galaxy}) is in essence the same expression used on page 173 of Carmeli \cite{Carmeli2002a} in his gravitational redshift formula rewritten here as
\begin{equation} \label{eqn:Cgravz}
\frac{\lambda_2}{\lambda_1} = \sqrt{\frac{1+ \Omega_K r^2_2/c^2\tau^2-R_S/r_2}{1+\Omega_K r^2_1/c^2\tau^2-R_S/r_1}}.
\end{equation}
involving a cosmological contribution ($\Omega_K r^2/c^2\tau^2$) and $R_S=2GM/c^2$, a local contribution where the mass $M$ is that of a compact object. The curvature ($\Omega_K$) results from the averaged mass/energy density of the whole cosmos, which determines how the galaxies `move' but motions of particles within galaxies is dominated by the mass  of the galaxy and the masses of the compact objects within. Therefore when considering the gravitational redshifts due to compact objects we can neglect any cosmological effects, only the usual Schwarzschild radius of the object need be considered. The cosmological contributions in (\ref{eqn:Cgravz}) are generally negligible. This then leads back to the realm of general relativity.

\section{\label{sec:WhiteHole}White Hole}

Now if we assume the radial extent of a finite matter distribution at the current epoch is equal to the current epoch scale radius, we can write
\begin{equation} \label{eqn:delta}
\Delta =\frac{1}{\sqrt{\Omega_k}} = \frac{1}{\sqrt{|1-\Omega_m|}},
\end{equation}
expressed in units of $c \tau$. In such a case, $\Delta = 1.02 \;c \tau$ if $\Omega_m = 0.04$ and $\Delta = 1.01 \; c \tau$ if $\Omega_m = 0.02$. 

It is important to note also that in Carmeli's unbounded model (\ref {eqn:phasespacesolnnatural})  describes the redshift distance relationship but there is no central potential. In Hartnett \cite{Hartnett2005} and in Oliveira and Hartnett \cite{Oliveira2006} equation (\ref {eqn:phasespacesolnnatural}) was curve fitted to the SNe Ia data and was found to agree with $\Omega_m = 0.02-0.04$ without the inclusion of dark matter or dark energy. Therefore the same conclusion must also apply to the finite bounded model suggested here.

In order to achieve a fit to the data, using either the finite bounded or unbounded models, the white hole solution of  (\ref{eqn:soln1field3}) or (\ref{eqn:soln1fieldgen2})  must be chosen. The sign of the terms in (\ref{eqn:fvalue1})  means that the potential implicit in (\ref{eqn:fvalue1}) is a potential hill, not a potential well. Therefore the solution describes an expanding white hole with the observer at the origin of the coordinates, the unique center of the Universe. Only philosophically can this solution be rejected. Using the Carmeli theory, the observational data cannot distinguish between finite bounded models ($\infty > \Delta \geq c\tau$) and finite ($\Delta = 0$) or infinite ($\Delta = \infty$) unbounded models  .

The physical meaning is that the solution, developed in this paper, represents an expanding white hole centered on the Galaxy. The galaxies in the Universe are spherically symmetrically distributed around the Galaxy. The observed redshifts are the result of cosmological expansion alone.

Moreover if we assume $\Delta \approx c \tau$ and $\Omega_m = 0.04$ then it can be shown \cite{Oliveira2006} that the Schwarzschild radius for the finite Universe
\begin{equation} \label{eqn:Rs}
R_s \approx \Omega_m \Delta = 0.04 \; c \tau.
\end{equation}
Therefore for a finite universe with $\Delta \approx c \tau$ it follows that $R_s \approx 0.04 \; c \tau \approx 200 \; Mpc$. Therefore an expanding finite bounded universe can be considered to be a white hole. As it expands the matter enclosed within the Schwarzschild radius gets less and less. The gravitational radius of that matter therefore shrinks towards the Earth at the center. 

This is similar to the theoretical result obtained by Smoller and Temple \cite{Smoller2002} who constructed a new cosmology from the FRW metric but with a shock wave causing a time reversal white hole. In their model the total mass behind the shock decreases as the shock wave expands, which is spherically symmetrically centered on the Galaxy. Their paper states in part ``...the entropy condition implies that the shock wave must weaken to the point where it settles down to an Oppenheimer Snyder interface, (bounding a finite total mass), that eventually emerges from the white hole event horizon of an ambient Schwarzschild spacetime.'' 

This result then implies that the earth or at least the Galaxy is in fact close to the physical center of the Universe. Smoller and Temple state \cite{Smoller2003} that ``With a shock wave present, the \textit{Copernican Principle is violated} in the sense that the earth then has a special position relative to the shock wave. But of course, in these shock wave refinements of the FRW metric, there is a spacetime on the other side of the shock wave, beyond the galaxies,
and so the scale of uniformity of the FRW metric, the scale on which the density of the galaxies is uniform, is no longer the largest length scale''[emphasis added]. 

Their shock wave refinement of a critically expanding FRW metric leads to a big bang universe of finite total mass. This model presented here also has a finite total mass and is a spatially flat universe. It describes a finite bounded white hole that started expanding at some time in the past. 

\section{\label{sec:Conclusion}Conclusion}

Since the Carmeli theory has been successfully analyzed with distance modulus data derived by the high-z type Ia supernova teams it must also be consistent with a universe that places the Galaxy at the center of an spherically symmetrical isotropic expanding white hole of finite radius. The result describes particles moving in both a central potential and an accelerating spherically expanding universe without the need for the inclusion of dark matter. The data cannot be used to exclude models with finite extensions $\Delta \geq c\tau$.

\end{document}